\documentclass{article}
\usepackage{spconf,amsmath,graphicx,hyperref}
\usepackage{amssymb}
\usepackage[capitalize,nameinlink]{cleveref}
\crefname{figure}{Fig.}{Figs.}
\usepackage{booktabs}
\usepackage{todonotes}
\usepackage[T1]{fontenc}

\title{TASK-ORIENTED NEURAL FOA ENCODING FOR SELD FROM IRREGULAR MICROPHONE ARRAYS}

\name{Jiachen Liu$^{1,2}$, Yin Cao$^{1,2}$, Ming Wu$^{1,2,3}$, Jun Yang$^{1,2,3}$}

\address{
$^{1}$Institute of Acoustics, Chinese Academy of Sciences, Beijing 100190, China\\
$^{2}$University of Chinese Academy of Sciences, Beijing 100049, China\\
$^{3}$State Key Laboratory of Acoustics and Marine Information, Beijing 100190, China
}
\begin{document}
\ninept
\maketitle
\begin{abstract}
Sound event localization and detection (SELD) systems often rely on first-order Ambisonics (FOA) input, whereas obtaining useful FOA representations from irregular microphone arrays remains challenging. This paper proposes a two-stage SELD framework that learns a task-oriented, FOA-compatible representation from microphone-array signals. A neural residual encoder first refines conventional FOA encoding through a signal-dependent correction. A teacher--student scheme then transfers event and spatial knowledge from theoretical FOA representations through frame-level permutation-invariant knowledge distillation. Experiments on synthetic scenes with tetrahedral and 12-channel Benchmark arrays, together with real stationary-source recordings from the LOCATA dataset, show that teacher guidance consistently improves downstream SELD performance and substantially reduces localization error. Signal-level analysis further shows that lower FOA reconstruction error does not necessarily correspond to better SELD performance, indicating that the distilled representation is optimized primarily for task-relevant spatial information rather than strict FOA reconstruction.
\end{abstract}
\section{Introduction}
\label{sec:intro}

Sound event localization and detection (SELD) jointly performs sound event detection (SED) and direction of arrival (DOA) estimation. It is an enabling technology for robot audition, acoustic surveillance, human--computer interaction, and immersive audio. Deep learning has substantially improved SELD performance under noise, reverberation, and overlapping sound events \cite{yeowEnvironmentalAcousticIntelligence2025, huSelectiveMemoryMetaLearningEnvironment2024}. However, most existing systems assume a specific spatial audio format or well-calibrated microphone geometry, which limits their applicability to irregular microphone arrays where accurate spatial encoding becomes challenging \cite{adavanneSoundEventLocalization2019, grumiauxSurveySoundSource2022}.

Existing SELD systems commonly operate on either first-order Ambisonics (FOA) signals or multichannel microphone signals. FOA-based methods typically combine log-Mel spectra with intensity vectors, which encode directional energy flow through the amplitude and phase relationships among spherical harmonic (SH) channels \cite{huSoundEventLocalization2022, nguyenSALSASpatialCueAugmented2022}. Microphone-domain methods instead use spatial features such as GCC-PHAT to capture inter-microphone time-delay information \cite{wangFourStageDataAugmentation2023, caoPolyphonicSoundEvent2019a}. Although microphone-domain features preserve array-specific spatial cues, their dimensions and spatial patterns are closely coupled with the number, ordering, and geometry of the microphones \cite{nagatomoWearableSeldDataset2022, kowalkGeometryAwareDOAEstimation2023}.

FOA provides a fixed-dimensional SH representation and can therefore reduce the dependence of the downstream SELD network on array geometry. However, obtaining accurate FOA signals requires an encoder matched to the array manifold. Conventional analytical and least-squares encoders perform well for suitably sampled regular arrays, but can be affected by spatial aliasing, ill-conditioned encoding matrices, and array-model mismatch for sparse or non-spherical arrays \cite{heikkinenNeuralAmbisonicsEncoding2024, gayerAmbisonicsEncodingArbitrary2024a}.

Recent neural Ambisonics encoders map microphone signals to SH signals and may incorporate microphone coordinates to accommodate array geometries \cite{heikkinenNeuralAmbisonicsEncoding2024, heikkinenGenAGeneralizingAmbisonics2025}. Residual approaches retain a linear encoder and learn corrective terms using a neural network \cite{deppischResidualLearningNeural2026}. However, these methods are primarily optimized with waveform, spectral, or SH reconstruction losses. Such objectives may not preserve event-discriminative and directional information required by multisource SELD. Moreover, direct SELD supervision on estimated FOA features may be insufficient because of the representation gap between theoretical and encoded FOA domains. This motivates teacher--student knowledge distillation \cite{hintonDistillingKnowledgeNeural2015} to transfer event and spatial information from theoretical FOA features.

In this work, we propose a two-stage SELD framework that learns a task-oriented FOA-compatible representation from microphone array signals. The main contributions are twofold. First, we develop a residual encoder initialized by conventional FOA encoding and jointly optimized with the downstream SELD network, providing a fixed four-channel interface for SELD with practical
microphone array geometries. Second, we introduce a teacher-guided learning scheme that uses theoretical FOA as privileged spatial information to transfer event and localization knowledge through frame-level permutation-invariant distillation. Unlike reconstruction-oriented Ambisonics encoding, the proposed framework optimizes the intermediate representation according to downstream SELD utility. Experiments on synthetic and real recordings show that teacher guidance consistently improves SELD performance over the unguided neural encoder and analytical encoding baselines.

\section{FOA Representation and Conventional Encoding}
\label{sec:foa_representation}

In this work, FOA adopts the ACN/SN3D convention \cite{nachbar2011ambix}, with FOA
channels ordered as $(W,Y,Z,X)$. For a sound field containing $Q$
sources, the theoretical FOA signal is given by \cite{adavanneMultiroomReverberantDataset2019a}
\begin{equation}
\label{eq:foa_signal}
\mathbf{a}(t)=
\sum_{q=1}^{Q}
\begin{bmatrix}
1 \\
\sin\phi_q\cos\theta_q \\
\sin\theta_q \\
\cos\phi_q\cos\theta_q
\end{bmatrix}
s_q(t),
\end{equation}
where $\phi_q$ and $\theta_q$ denote the azimuth and elevation of the
$q$-th source, respectively, and $s_q(t)$ denotes its waveform.
Since this representation is directly constructed from the source
signals and DOAs without array-induced encoding errors, it serves as
the theoretical reference and the privileged input to the teacher
SELD network.


Let $\mathbf{X}(n,k)=[X_1(n,k),\ldots,X_M(n,k)]^{T}$ denote the STFT coefficients of an $M$-microphone array. Conventional FOA encoding is based on spherical modal-domain inversion \cite{rafaely2015fundamentals}. For a non-spherical array, we approximate the geometry by a virtual sphere with mean radius $\bar r=\sum_{m=1}^{M} r_m/M$, while preserving the microphone directions. The encoded FOA signal is
\begin{equation}
    \mathbf{A}_{\mathrm{conv}}(n,k)
    =\mathbf{B}^{-1}(k;\bar r)
    \mathbf{Y}_{\mathrm{sph}}^{\dagger}
    \mathbf{X}(n,k),
    \label{eq:conventional_foa}
\end{equation}
where $\mathbf{Y}_{\mathrm{sph}}$ is the first-order SH sampling matrix and $\mathbf{B}(k;\bar r)$ is the corresponding mode-strength matrix. This spherical approximation introduces model mismatch and provides the initialization for the proposed residual encoder.

\section{PROPOSED TWO-STAGE SELD FRAMEWORK}
\label{sec:proposed}

\begin{figure*}[t]
	\centering
        \includegraphics[width=\linewidth]{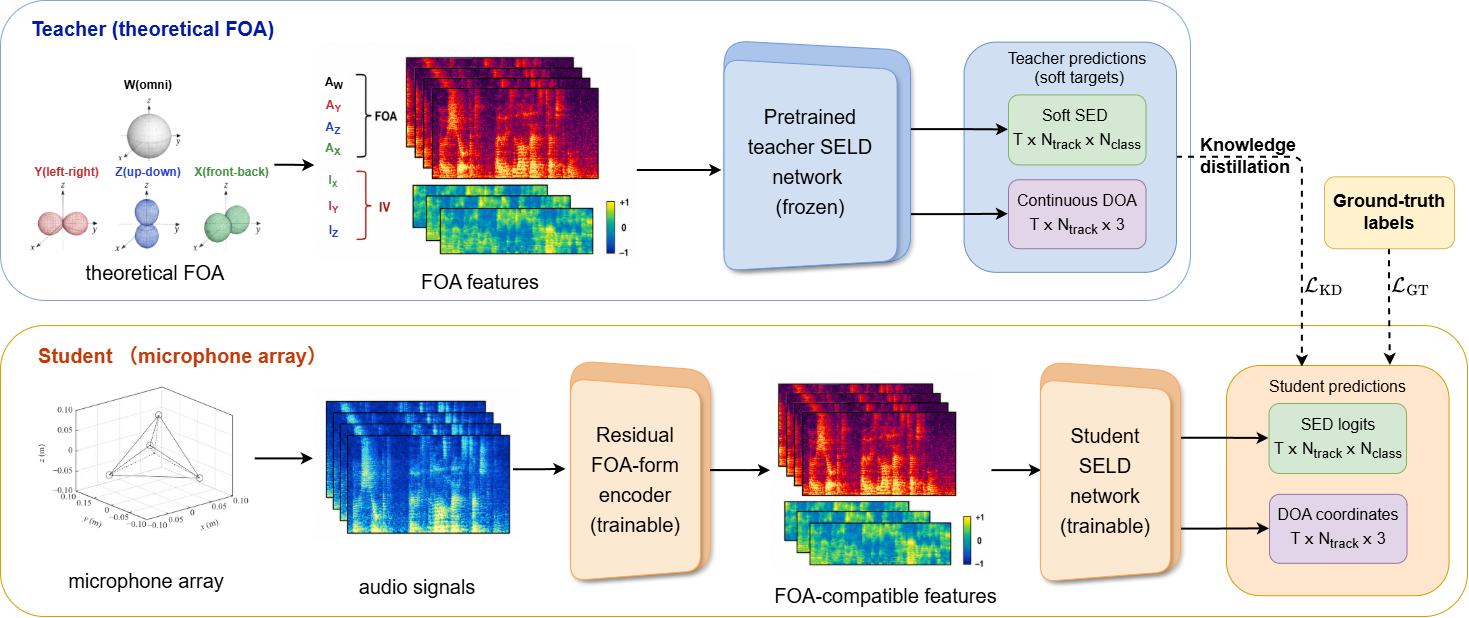}
	  \caption{Overview of the proposed teacher--student SELD framework}\label{network}
\end{figure*}

The proposed framework aims to enable SELD from an irregular microphone array by learning a task-oriented, FOA-compatible spatial representation. Rather than explicitly optimizing the intermediate representation for exact FOA coefficient reconstruction, the proposed framework jointly optimizes the neural encoder and the downstream SELD network under guidance from a theoretical-FOA teacher. As illustrated in \cref{network}, the framework consists of a trainable residual encoder, a student SELD network, and a pretrained teacher SELD network.

\subsection{Residual FOA-Form Encoder}
\label{ssec:foa_encoder}

Given the microphone-domain STFT coefficients $\mathbf{X}(n,k)\in\mathbb{C}^{M}$, the residual encoder produces a four-channel FOA-compatible representation as \cite{deppischResidualLearningNeural2026}
\begin{equation}
\label{eq:neural_foa_encoder}
\widehat{\mathbf{A}}(n,k) = \left[\mathbf{E}(k) + \Delta\mathbf{E}_{\boldsymbol{\varphi}} \bigl(n,k;\mathbf{X}\bigr) \right]\mathbf{X}(n,k),
\end{equation}
where $\mathbf{E}(k)=\mathbf{B}^{-1}(k;\bar r)\mathbf{Y}^{\dagger}_{\mathrm{sph}}$ is the conventional encoder in \cref{eq:conventional_foa}, and $\Delta\mathbf{E}_{\boldsymbol{\varphi}} (n,k;\mathbf{X})\in\mathbb{C}^{4\times M}$ is a signal-dependent residual encoding matrix predicted by the neural network. The analytical encoder provides a physically motivated initial estimate, whereas the learned residual can compensate for array-response mismatch and unmodeled magnitude and phase distortions.

The encoder output $\widehat{\mathbf{A}}(n,k)$ follows the same four-channel $(W,Y,Z,X)$ organization and time--frequency structure as FOA and is initialized from the conventional FOA mapping. However, because the residual encoder is jointly optimized for the downstream SELD objective without an explicit coefficient-level FOA reconstruction constraint, $\widehat{\mathbf{A}}(n,k)$ is treated as an FOA-compatible task-oriented representation rather than an exact estimate of the theoretical FOA coefficients.

The network takes the concatenated real and imaginary STFT components as input and predicts the complex-valued residual encoding matrix. 

\subsection{Teacher--Student SELD Learning}
\label{ssec:teacher_student}

Direct SELD supervision can adapt the intermediate representation toward task-specific information. Rather than constraining this representation to exactly reconstruct the theoretical FOA coefficients, we exploit theoretical FOA as privileged information through teacher--student knowledge distillation \cite{hintonDistillingKnowledgeNeural2015}. The teacher therefore provides event- and localization-level guidance, allowing the encoder to emphasize spatial information useful for SELD while retaining the four-channel FOA-compatible interface.

Both the theoretical FOA and the FOA-compatible representation are converted into the same seven-channel SELD representation,, consisting of four channel-wise log-Mel spectrograms and three normalized active-intensity components \cite{huSoundEventLocalization2022}. The corresponding features are denoted by $\mathbf{F}$ and $\widehat{\mathbf{F}}$, respectively.


The teacher and student adopt the same track-wise SELD architecture \cite{huSoundEventLocalization2022}, but have independent parameters and receive different inputs. The pretrained and frozen teacher processes $\mathbf{F}$ extracted from the clean theoretical FOA defined in \cref{eq:foa_signal}, whereas the student processes $\widehat{\mathbf{F}}$ extracted from the FOA-compatible representation derived from noisy room-rendered microphone signals. Thus, the theoretical-FOA teacher provides privileged task-level supervision rather than a signal-level reconstruction target. The teacher remains frozen, while the student and residual encoder are jointly optimized through ground-truth supervision and knowledge distillation.

\subsection{Training Objective}
\label{ssec:training_objective}

The student is supervised by the ground-truth SELD labels using the frame-level tPIT loss adopted in \cite{huSoundEventLocalization2022}. Binary cross-entropy and mean squared error are used for SED and DOA estimation, respectively, with the optimal track permutation selected independently for each frame.

For knowledge distillation, frame-level tPIT is also applied between the student and teacher outputs. For each candidate teacher-track permutation $\pi$, the SED distillation loss is defined as

\begin{equation}
\ell_{\mathrm{KD},\pi}^{\mathrm{SED}} = \operatorname{BCE}_{\mathrm{logit}} \left(\mathbf{Z}^{\mathrm{S}}, \sigma\mathbf{Z}^{\mathrm{T}}_{\pi} \right),
\label{eq:kd_sed}
\end{equation}
where $\mathbf{Z}^{\mathrm{S}}$ and $\mathbf{Z}_{\pi}^{\mathrm{T}}$ denote the track-wise SED logits of the student and permuted teacher, respectively. $\sigma(\cdot)$ denotes the sigmoid function. $\operatorname{BCE}_{\mathrm{logit}}(\cdot,\cdot)$ denotes binary cross-entropy computed directly from logits.

The confidence assigned to student track $j$ under a candidate teacher-track permutation $\pi$ is defined as
\begin{equation}
c_{\pi,j} = \max_q \sigma\left( Z_{\pi(j),q}^{\mathrm{T}} \right),
\end{equation}
where $\pi(j)$ denotes the teacher track matched to student track $j$, and $q$ indexes the sound-event classes. The corresponding confidence-weighted DOA distillation loss is \begin{equation}
\ell_{\mathrm{KD},\pi}^{\mathrm{DOA}} = \frac{ 
 \sum_j c_{\pi,j} \mathbb{I}(c_{\pi,j}\geq\eta) 
\left\| \mathbf{D}_{j}^{\mathrm{S}} - \mathbf{D}_{\pi(j)}^{\mathrm{T}} \right\|_2^2
}{ \sum_j c_{\pi,j} \mathbb{I}(c_{\pi,j}\geq\eta) +\epsilon
}.
\label{eq:kd_doa_loss}
\end{equation}
where $\eta$ is the teacher-confidence threshold and $\mathbb{I}(\cdot)$ denotes the indicator function. The permutation minimizing the weighted sum of $\ell_{\mathrm{KD},\pi}^{\mathrm{SED}}$ and
$\ell_{\mathrm{KD},\pi}^{\mathrm{DOA}}$ is selected independently for each frame. 
The overall objective is
\begin{equation}
\label{eq:total_loss}
\mathcal{L}=\sum_{u\in\{\mathrm{GT},\mathrm{KD}\}}\lambda_u
\left[\beta_u\mathcal{L}_u^{\mathrm{SED}}+(1-\beta_u)\mathcal{L}_u^{\mathrm{DOA}}\right].
\end{equation}
We set $\beta_{\mathrm{GT}}=\beta_{\mathrm{KD}}=0.5$, $\eta=0.5$, $\lambda_{\mathrm{GT}}=1$, and $\lambda_{\mathrm{KD}}=0.2$. Gradients from both objectives are propagated through the student and residual encoder, guiding the intermediate representation toward the event and directional information conveyed by the theoretical-FOA teacher.

\begin{table*}[t]
\centering
\caption{SELD performance on noisy synthetic microphone-array
data. Theoretical FOA is shown as a common oracle reference.
Boldface denotes the best practical FOA-based result in each condition.}
\label{tab:synthetic}
\resizebox{\linewidth}{!}{
\begin{tabular}{l|cccc|c|cccc|c}
\toprule
& \multicolumn{5}{c|}{Tetrahedral Array}
& \multicolumn{5}{c}{Benchmark Array} \\
\cmidrule(lr){2-6}
\cmidrule(lr){7-11}
Method
& ER$_{20^\circ}\downarrow$
& F$_{20^\circ}\uparrow$
& LE$_{CD}\downarrow$
& LR$_{CD}\uparrow$
& $\mathcal{E}_{SELD}\downarrow$
& ER$_{20^\circ}\downarrow$
& F$_{20^\circ}\uparrow$
& LE$_{CD}\downarrow$
& LR$_{CD}\uparrow$
& $\mathcal{E}_{SELD}\downarrow$ \\
\midrule

Theoretical FOA
& 0.456 & 60.7\% & 9.3$^\circ$ & 61.7\% & 0.321
& 0.436 & 62.9\% & 8.4$^\circ$ & 62.7\% & 0.307 \\







Conventional FOA
& 1.092 & 10.3\% & 60.4$^\circ$ & 40.4\% & 0.730
& 0.699 & 40.1\% & 30.7$^\circ$ & \textbf{52.8\%} & 0.485 \\

ASM
& 1.012 & 9.5\% & 77.4$^\circ$ & 43.9\% & 0.727
& 0.970 & 18.3\% & 62.3$^\circ$ & 34.3\% & 0.698 \\

Neural FOA
& 0.809 & 30.9\% & 26.8$^\circ$ & 44.5\% & 0.551
& 0.780 & 29.6\% & 34.3$^\circ$ & 47.7\% & 0.549 \\

Neural FOA + KD
& \textbf{0.641} & \textbf{45.0\%} & \textbf{17.0$^\circ$}
& \textbf{49.8\%} & \textbf{0.447}
& \textbf{0.611} & \textbf{47.1\%} & \textbf{15.1$^\circ$}
& 51.0\% & \textbf{0.429} \\

\bottomrule
\end{tabular}}
\end{table*}

\begin{table}[t]
\centering
\caption{Signal-level fidelity of intermediate representations on noisy synthetic data. The theoretical FOA is used as the reference.}
\label{tab:foa_matrix}
\resizebox{\linewidth}{!}{
\begin{tabular}{l|cc|cc}
\toprule
& \multicolumn{2}{c|}{Tetrahedral Array}
& \multicolumn{2}{c}{Benchmark Array} \\
\cmidrule(lr){2-3}
\cmidrule(lr){4-5}
Method
& NMSE (dB) $\downarrow$
& IV-MSE $\downarrow$
& NMSE (dB) $\downarrow$
& IV-MSE $\downarrow$ \\
\midrule

Conventional FOA
& 2.301 & 0.280
& 2.057 & 0.268 \\

ASM
& \textbf{1.031} & \textbf{0.258}
& \textbf{1.496} & \textbf{0.240} \\

Neural FOA
& 1.380 & 0.275
& 1.927 & 0.264 \\

Neural FOA + KD
& 2.814 & 0.266
& 3.511 & 0.287 \\

\bottomrule
\end{tabular}
}
\end{table}

\section{EXPERIMENTS AND RESULTS}
\label{sec:EXPERIMENTS}

\begin{table*}[t]
\centering
\caption{SELD performance comparison on the LOCATA Benchmark dataset. Boldface denotes the best practical FOA-based result.}
\label{tab:locata_results}
\resizebox{\linewidth}{!}{
\begin{tabular}{l|cccc|c|cccc|c}
\toprule
& \multicolumn{5}{c|}{Task 1 (Single Source)}
& \multicolumn{5}{c}{Task 2 (Multi Source)}\\
\cmidrule(lr){2-6}
\cmidrule(lr){7-11}
Method 
& ER$_{20^\circ}\downarrow$ 
& F$_{20^\circ}\uparrow$ 
& LE$_{CD}\downarrow$ 
& LR$_{CD}\uparrow$ 
& $\mathcal{E}_{SELD}\downarrow$
& ER$_{20^\circ}\downarrow$ 
& F$_{20^\circ}\uparrow$ 
& LE$_{CD}\downarrow$ 
& LR$_{CD}\uparrow$ 
& $\mathcal{E}_{SELD}\downarrow$ \\
\midrule




Theoretical FOA
& 0.465 & 55.8\% & 3.2$^\circ$ & 56.3\% & 0.340
& 0.751 & 31.4\% & 14.5$^\circ$ & 34.7\% & 0.543 \\

Conventional FOA
& 1.121 & 7.1\% & 40.6$^\circ$ & 39.5\% & 0.720
& 1.009 & 0.0\% & 93.6$^\circ$ & 17.1\% & 0.840 \\

ASM
& 0.984 & 23.9\% & 26.3$^\circ$ & 34.8\% & 0.636
& 0.900 & 14.6\% & 53.9$^\circ$ & \textbf{30.4\%} & 0.690 \\

Neural FOA + KD
& \textbf{0.820} & \textbf{26.1\%} & \textbf{23.6$^\circ$} & \textbf{46.9\%} & \textbf{0.555}
& \textbf{0.868} & \textbf{18.6\%} & \textbf{41.0$^\circ$} & 28.5\% & \textbf{0.656} \\

\bottomrule
\end{tabular}
}
\end{table*}

\subsection{Datasets}
\label{ssec:Datasets}

The simulated dataset was constructed using the 12 sound classes specified in the official \texttt{FSD50K\_selected.txt} file for DCASE 2022 Task 3 \cite{politisDatasetDynamicReverberant2021a, fonsecaFSD50KOpenDataset2022}. We selected 200 recordings from each class to construct the simulated scenes. For each class, 90\% of the recordings were used to generate the training scenes, while the remaining 10\% were reserved for generating the validation and test scenes, ensuring that no source recording was shared between the training and evaluation sets. Each 10-s scene contained one to three stationary sources and was rendered using Pyroomacoustics \cite{scheiblerPyroomacousticsPythonPackage2018} for a four-channel tetrahedral array with a radius of $0.1$~m and the 12-channel LOCATA Benchmark array \cite{eversLOCATAChallengeAcoustic2020}. Room dimensions were randomly sampled from $[4,20]\times[4,20]\times[3,10]$~m, with a wall absorption coefficient of 0.99. For each scene, clean microphone-array signals were first generated, and the corresponding noisy signals were obtained by adding white noise at an SNR uniformly sampled from 6 to 30~dB while keeping the source and room configurations unchanged. The theoretical FOA signals were generated using the same source signals and DOAs as the corresponding microphone-array recordings. The training, validation, and test sets contained 20,000, 2,000, and 2,000 scenes, respectively.

For real-data evaluation, we use the stationary-source recordings from the LOCATA dataset captured by the 12-channel Benchmark array. Specifically, all 16 Task 1 recordings are included, while Task 2 recordings with four simultaneous sources are excluded, resulting in 11 multi-source recordings for evaluation. These recordings contain stationary sources and a stationary array. Since LOCATA does not provide sound-event class annotations, the speech categories are manually labeled as male or female speech based on the reference source signals. Theoretical FOA references are synthesized from the reference source signals and ground-truth directions provided by LOCATA. 

\subsection{Implementation Details}\label{ssec:details}

All signals were sampled at 16~kHz. STFT was computed using a 510-sample Hann window and a 200-sample hop, producing 256 single-sided frequency bins. Each input segment contained 128 frames (1.62~s). For an $M$-channel array, the real and imaginary STFT components were concatenated into a $2M\times128\times256$ tensor, yielding 8 and 24 input channels for the tetrahedral and Benchmark arrays, respectively.

The residual network contains six ConvGLU encoder blocks and six DeConvGLU decoder blocks \cite{tanLearningComplexSpectral2020}. A pair of two stage Mamba (TS-Mamba) blocks \cite{wangMambaSEUNetMambaUNet2024} are used at the bottleneck, followed by another two TS-Mamba blocks and a linear mapping layer for residual estimation. Each bidirectional Mamba module uses $d_{\mathrm{state}}=16$, $d_{\mathrm{conv}}=4$, an expansion factor of 4, and RMS normalization. The teacher and student use the track-wise SELD architecture in \cite{huSoundEventLocalization2022}, with three output tracks. The simulated-data models use 12 classes, whereas the LOCATA models use two speech classes.

The teacher was pretrained on theoretical FOA features and then frozen. The FOA encoder and student network were jointly trained using Adam with a learning rate of $0.0001$ and a batch size of 32, for up to 100 epochs with an early-stopping patience of 10 epochs. Separate models were trained for the two array configurations.

\subsection{Evaluation on Synthetic Data}
\label{ssec:synthetic_results}

Table~\ref{tab:synthetic} compares the proposed systems with analytical FOA baselines on both arrays. Besides Conventional FOA, we evaluate Ambisonics signal matching (ASM) \cite{gayerAmbisonicsEncodingArbitrary2024a}, a regularized linear encoder that matches array steering responses to desired SH responses. Theoretical and Conventional FOA are obtained using \cref{eq:foa_signal,eq:conventional_foa}, respectively. Each analytical representation is evaluated with an independently trained SELD network. Neural FOA jointly optimizes the residual encoder and SELD network using ground-truth supervision, while Neural FOA + KD additionally uses the frozen theoretical-FOA teacher. All systems share the same SELD architecture and training settings. SELD performance follows the DCASE metrics \cite{politisDatasetDynamicReverberant2021a}. Signal-level fidelity is measured by overall NMSE relative to theoretical FOA and IV-MSE between normalized active-intensity components.

On the tetrahedral array, Neural FOA consistently outperforms the analytical encoding baselines, while on the Benchmark array it remains inferior to Conventional FOA in overall SELD performance but performs better than ASM. Introducing KD further improves Neural FOA on both array configurations, with particularly clear gains in event detection and localization accuracy. The distilled model achieves the best overall SELD performance among the practical methods, although it does not consistently obtain the highest localization recall. These results show that teacher guidance effectively steers the FOA-compatible intermediate representation toward event and spatial information that is more useful for downstream SELD.

Table~\ref{tab:foa_matrix} further evaluates the signal-level fidelity of the intermediate representations using theoretical FOA as the reference. Without distillation, the neural residual encoder improves the overall NMSE over Conventional FOA on both arrays, indicating that residual refinement can enhance coefficient-level fidelity. ASM achieves the best signal-level reconstruction quality overall, whereas the representation learned with KD exhibits higher NMSE and no consistent improvement in IV-MSE. However, this trend differs markedly from the downstream SELD results in Table~\ref{tab:synthetic}, where the distilled model achieves the best overall performance among the practical methods. This discrepancy indicates that signal-level FOA fidelity alone does not determine the downstream utility of the learned representation.

\subsection{Evaluation on Real Recordings}
\label{ssec:real}

Table~\ref{tab:locata_results} reports the results on the LOCATA Benchmark dataset under single-source (Task 1) and multi-source (Task 2) scenarios. ASM generally improves over Conventional FOA by incorporating measured array responses, demonstrating the benefit of calibration-based encoding under real-array mismatch. However, its performance remains substantially below that of theoretical FOA, indicating that analytical encoding remains sensitive to practical array-response variations.

Neural FOA + KD achieves the best overall performance among the practical methods in the single-source scenario, with advantages in both event detection and localization over the analytical baselines. In the multi-source scenario, it achieves lower ER and localization error and a better overall SELD score than both analytical encoders, although ASM obtains slightly higher localization recall. These results show that the proposed task-oriented FOA-compatible representation provides more effective spatial information for SELD under real recording conditions.

\section{CONCLUSION}
\label{sec:conclusion}

This paper proposed a two-stage SELD framework that learns a task-oriented, FOA-compatible spatial representation from microphone-array signals. A neural residual encoder builds on conventional FOA encoding, while a teacher--student scheme transfers event and directional knowledge from theoretical FOA through frame-level permutation-invariant distillation. Experiments on noisy synthetic data show that teacher guidance consistently improves SELD performance and substantially reduces localization error on both tetrahedral and Benchmark arrays. Signal-level analysis further reveals that the lowest FOA reconstruction error does not necessarily yield the best downstream SELD performance, highlighting the distinction between coefficient-level fidelity and task-oriented spatial representation quality. Experiments on real LOCATA recordings further demonstrate improved SELD performance over analytical encoding baselines in both single- and multi-source scenarios. These results support FOA-compatible task-oriented representation learning as an effective interface between microphone arrays and FOA-based SELD systems.

\bibliographystyle{IEEEbib}
\bibliography{bib}

\end{document}